# Physics-grounded generative design of inherently stable, novel and controllable crystal structures

Zhilong Song[1], Qionghua Zhou[1,2,*], Chongyi Ling[1], Qiang Li[1,2], Lixue Cheng[3*] and Jinlan Wang[1,2,*]

[1]Key Laboratory of Quantum Materials and Devices of Ministry of Education, School of Physics, Southeast University, Nanjing 21189, China

[2] Suzhou Laboratory, Suzhou 215000, China

[3] Department of Chemistry, Hong Kong University of Science and Technology, Hong Kong 999077, China

Generative inverse design is reshaping the discovery of functional crystalline materials. Yet current generative models face challenges in simultaneously achieving stability, novelty, and precise controllability in a single trained model. We address these challenges with a key physical insight: the diversity of crystals is governed by their crystallographic information (CI), namely composition, space group and lattice, whereas only a few stable atomic configurations remain once the CI is fixed. Built on this insight, we introduce SCGEN (stable and controllable crystal structure generation), a physics-grounded generative model with two components. A variational autoencoder samples diverse, physically plausible CI, and a symmetry- and Wyckoff-position-constrained optimizer locates stable atomic positions via universal machine-learning potentials. Benchmarked on roughly two million structures, SCGEN reaches state-of-the-art stability while preserving comparable novelty, and it satisfies any specified composition, space group, lattice or joint constraint with 100% success and no task-specific retraining. Applied to photocatalytic water splitting, property-guided optimization with SCGEN generates 200,000 candidate structures and identify top 22 stable, active and synthesizable photocatalysts. By decoupling CI generation from coordinate optimization, SCGEN establishes a physics-grounded inverse-design paradigm that yields synthesis-ready crystals on demand, rather than structures requiring post hoc repair, relaxation, or retraining.

## 1. Introduction

Machine learning is progressively accelerating the discovery of functional materials[1–6]. In the forward design paradigm, learned models map composition or structure to properties and accelerate the screening of known compounds. The harder inverse paradigm starts from a target property and produces the corresponding crystal, including material structures never synthesized, and has become tractable only recently through generative models[7,8]. Early demonstrations based on variational autoencoders (VAEs) and generative adversarial networks (GANs) were limited to specific material families, such as V-O[9] systems, Bi-Se[10] systems, Mg−Mn−O[11] systems, zeolites[12], cubic semiconductors[13] and metal–organic frameworks[14]. Diffusion- and flow-based models then overcame this limitation: CDVAE[15], DiffCSP++[16], MatterGen[17], FlowMM[18] and CrystalFlow[19] generate diverse bulk and two-dimensional crystals by denoising or transporting samples from a prior distribution to the data manifold while respecting crystallographic invariances. Going further, SCIGEN[20] integrated hard geometric-lattice constraints into a pretrained diffusion generator without retraining, steering generation toward candidate quantum materials at the scale of ten million structures. Together, these architectures can sample novel chemistries and structural motifs that extend considerably beyond the training distribution.

However, these advances share a common limitation. Each applies generative mechanisms originally developed for computer vision and natural language directly to crystals, generating atomic coordinates and crystallographic information (CI, namely composition, space group and lattice) simultaneously while capturing the underlying physics, energetics and symmetry only implicitly from data. This implicit treatment has three limitations. First, because first-principles energetics never enter the sampling or denoising process, generated structures often lie far above the thermodynamic convex hull before density functional theory (DFT) relaxation. Second, without explicit symmetry constraints, the generated atoms rarely occupy their exact Wyckoff positions, causing the symmetry to degrade to the trivial P1 group. Third, because CI is produced jointly with the coordinates, the generator cannot be steered toward an arbitrary composition, space group and lattice without task-specific retraining. Even SCIGEN, the most explicit attempt to embed physical structure into a generator, constrains only a fixed motif library and recovers stability post hoc.

The solution is well established in other fields: embedding the physics directly, rather than learning it implicitly. Across the computational sciences, building governing equations[21], energy[22] or symmetry group[23] into the model itself delivers the data efficiency and physical reliability that purely learned models lack[21]. In crystals, this physics-grounded paradigm is embodied by crystal-structure prediction (CSP). Long-established engines such as CALYPSO[24] and USPEX[25] search the energy landscape under full crystallographic symmetry. Recently, universal machine-learning interatomic potentials (UMLP) such as MatterSim[26] and MACE[27], together with UMLP-native pipelines such as GNoME[28], have made this search tractable at database scale, directly locating stable or metastable structures via symmetry-constrained energy minimization. However, CSP optimizes only atomic coordinates within a user-supplied composition or space group, its chemical novelty depends entirely on the external pipeline that feeds it. Inverse materials design therefore confronts two complementary bottlenecks: generative models explore chemical space broadly but yield structures of unreliable stability due to lacking embedded physics, whereas CSP delivers physically grounded stability at the cost of novelty.

We bridge these two bottlenecks from a physical insight: the CI governs structural diversity, whereas once CI is fixed, atomic positions are restricted to a few Wyckoff degrees of freedom and the energy landscape is dominated by a handful of large-basin, high-symmetry minima[29,30]. Therefore, generating high-quality novel crystal structures does not require directly generating atomic coordinates. Based on this insight, we introduce SCGEN (Stable and Controllable crystal structure GENeration), a physics-grounded generative model that decouples the generation process into two complementary stages. A crystallographic information variational autoencoder (CIVAE) inherits the chemical novelty of generative models, sampling diverse, plausible CI. A stable coordinates optimizer (SCO) inherits the physical grounding of CSP, locating stable atomic positions on a UMLP energy landscape under strict Wyckoff and symmetry constraints. Because SCO accepts CI from either CIVAE sampling or direct user input, any composition, space group, lattice or combination can be enforced without task-specific retraining. On a unified benchmark of 1,971,023 inorganic crystals integrated from seven major repositories, SCGEN achieves the state-of-the-art average energy above hull (0.021 eV/atom)—outperforming MatterGen (0.036 eV/atom), while preserving comparable novelty and satisfying every individual or joint specification of composition, space group and lattice with 100% success. Applied to photocatalytic water splitting[31,32], property-guided optimization with SCGEN generates 200,000 candidates (81.2% novel), from which a hierarchical computational pipeline identifies 22 DFT-validated photocatalysts. By decoupling CI generation from coordinate optimization, SCGEN establishes a physics-grounded paradigm for inverse design in which stability, symmetry and controllability hold by construction and every structure is tailored to target specifications.

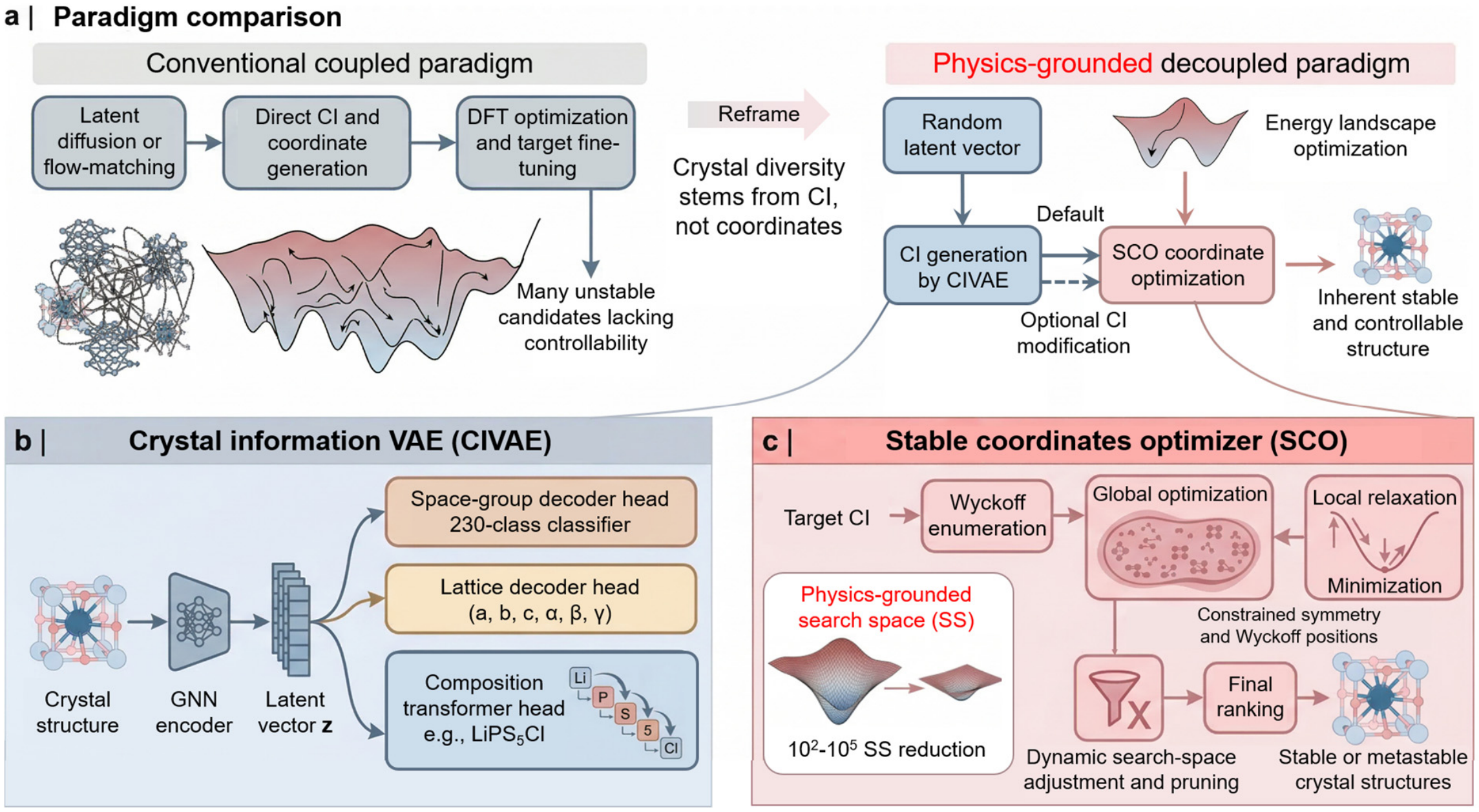

**Fig. 1. (a)** Paradigm comparison between the conventional coupled paradigm (left), which generates crystallographic information (CI) and coordinates together via latent diffusion or flow-matching followed by DFT optimization, often yielding unstable candidates lacking controllability, and the SCGEN physical-grounded decoupled paradigm (right), built on the principle that crystal diversity stems from CI rather than coordinates: a random latent vector is mapped to CI by the CIVAE (with optional CI modification), and the SCO then optimizes coordinates on the energy landscape to yield inherently stable, controllable structures. **(b)** Architecture of the crystallographic information VAE (CIVAE). A GNN encoder projects crystal structures into a continuous latent vector z; this symmetry-aware embedding enables stochastic sampling of novel crystallographic information. The decoder comprises three heads: a space-group classifier (230 classes), a lattice parameter regressor ($a, b, c, \alpha, \beta, \gamma$), and a Transformer-based composition decoder (e.g., $LiPS_5Cl$). **(c)** Workflow of the stable coordinates optimizer (SCO): target CI is processed by Wyckoff enumeration and physical-grounded search-space (SS) reduction ($10^2$–$10^5$-fold), hierarchical global optimization and local relaxation under constrained symmetry and Wyckoff positions, with dynamic pruning and final ranking to yield stable or metastable crystal structures.

## 2. Results and discussion

### 2.1 Architecture and module performance of SCGEN

The architecture of SCGEN follows physical principles rather than a purely data-driven design choice. It decouples generation into two modules that communicate solely through a discrete CI layer (Fig. 1a): the CIVAE samples CI from a continuous latent space, and the SCO converts each sample into a stable atomic configuration on a UMLP energy landscape. This decoupling yields two advantages. First, fixing atomic positions through physical optimization rather than a learned decoder keeps the network compact, since it need only model the low-dimensional CI, and makes stability and symmetry intrinsic rather than learned. Second, because the modules interact solely through the discrete CI, each can be upgraded independently without mutual retraining.

The CIVAE (Fig. 1b) treats composition, space group and lattice parameters as three decoupled generative targets. An equivariant graph neural network encoder (e.g., DimeNet++[33] or GemNet[34]) maps each input crystal into a continuous 256-dimensional latent vector **z** while preserving crystallographic symmetry. From this shared representation, three specialized decoder heads jointly reconstruct the full crystallographic description. The first is a fully connected classifier that projects z onto a probability distribution over all 230 space groups. The second is a fully connected regressor that predicts the six lattice parameters ($a$, $b$, $c$, $\alpha$, $\beta$, $\gamma$). The third is a Transformer-based autoregressive decoder that generates the chemical formula token by token, conditioned on the predicted space group and lattice parameters. Compositions use a 173-token vocabulary: 112 element tokens, 60 stoichiometric-multiplicity tokens and one termination token. For instance, $Cu_1Al_2$ is tokenized as [28, 112, 12, 113], where element-identity tokens (28 → Cu, 12 → Al) alternate with count tokens (112 → "1", 113 → "2"). At each step, the composition network receives a 492-dimensional vector concatenating the space-group probability vector (230 dimensions), the lattice parameters (6 dimensions) and the latent vector z (256 dimensions); this vector then passes through token-embedding, Transformer self-attention and linear projection layers, from which the network predicts the next token. This conditioning is essential, i.e., architectures that disrupt this conditioning scheme (e.g. alternative cascaded and parallel ones) can lead to higher validation losses (Table S2).

The SCO module (Fig. 1c) then turns these CIVAE outputs into thermodynamically stable atomic configurations by systematically exploring Wyckoff-position space. Given the predicted space group, lattice parameters and composition, SCO enumerates all valid Wyckoff-position assignments, then couples swarm-intelligence optimization (here the bird swarm algorithm[35]) with UMLPs such as MatterSim[26] and MACE[27] to identify the most thermodynamically stable configurations. The central obstacle is the combinatorial complexity of Wyckoff assignment—$Li_2FeO_3$ in space group 21, for example, admits $10^6$ distinct configurations. To make this combinatorial search tractable, we developed four complementary strategies that exploit the underlying crystallographic physics—symmetry, Wyckoff multiplicity and the true coordinate degrees of freedom—to confine the search to the physically meaningful subspace rather than the full coordinate space.

i) Accelerated Wyckoff-combination enumeration (Fig. S1). Valid combinations grow exponentially with chemical complexity and decreasing space-group symmetry, which is most severe for quaternary and quinary compositions in low-symmetry groups. A depth-first search prunes invalid branches before expansion, and a validity filter discards configurations violating Wyckoff multiplicity, site symmetry or chemical feasibility; survivors are combined by Cartesian product. Above 20,000 combinations a simplified enumeration is used, and parallelization yields a 5-fold speed-up (Supplementary Note S1).

ii) Hierarchical global–local optimization at fixed Wyckoff position and space group (Fig. S2). A purely global search ignores continuous relaxation within each assignment, reaching only −0.449 eV/atom on MP-20. We therefore nest a symmetry-constrained local relaxation inside it: the independent coordinates are read from the Wyckoff site symmetries and the rest generated from them via space-group operations, so relaxation moves only the true degrees of freedom while keeping the structure exactly symmetric (Fig. S3, Supplementary Note S2). By minimizing the UMLP potential energy via conjugate-gradient relaxation, we lower the mean predicted formation energy to −0.695 eV/atom (Fig. S4b, Table S1). We term this variant SCGEN-S.

iii) Primitive-cell-based Wyckoff optimization (Fig. S5). A conventional cell replicates symmetry-equivalent atoms with non-independent coordinates; mapping it to the primitive cell removes these duplicates—for example from 21 to 7 atoms—with no loss of crystallographic information. Because positions here are symbolic Wyckoff variables rather than numbers, the transformation to the primitive cell requires symbolic matrix algebra and constraint propagation, which we perform per candidate and distribute across processors (Supplementary Note S3).

iv) Dynamic search-space adaptation tied to the chosen Wyckoff positions (Fig. S6). The coordinate-search dimensionality is set by the Wyckoff assignment: a naive 21-atom structure carries 64 variables, most bound by symmetry. The optimizer identifies the genuinely independent coordinates and rebuilds the problem in real time at the matching dimension—collapsing, for example, 64 variables to seven z-coordinates. Matching the search dimension to the true degrees of freedom contracts the effective search space by 60–85% (Supplementary Note S4).

With both modules fully defined, we now evaluate their performances. Because CIVAE needs only learn the low-dimensional crystallographic information rather than the full atomic coordinate field, it faces a far better-posed inverse problem than conventional generators, and its reconstruction fidelity reflects this directly. On the held-out test set it recovers lattice lengths and angles at 96.1% and 98.8% accuracy, surpassing both MatterGen (93.5% and 97.6%) and CDVAE (92.8% and 95.4%) (Fig. 2a). It also reconstructs compositions at 95.6% accuracy with 95.4% chemical plausibility, versus 95.2% and 89.3% for MatterGen and 92.8% and 85.9% for CDVAE (Fig. 2b). In the space-group confusion matrix (Fig. 2c), diagonal accuracy exceeds 88% across every symmetry range, and the residual off-diagonal weight falls almost exclusively on crystallographically related groups rather than scattering at random. This indicates that the network has internalized the hierarchical symmetry relationships rather than memorizing labels, a consequence of conditioning the composition decoder on the predicted space group and lattice to enforce mutual self-consistency.

We next quantify how each SCO strategy improves the stability of generated structures. The guiding principle is that every strategy removes degrees of freedom that crystallographic symmetry has already fixed, ensuring that the optimizer spends its effort only on the coordinates that are genuinely free to move and converges closer to a true energy minimum. The improvement is monotonic: activated in sequence, the four strategies drive the mean predicted formation energy of randomly generated structures from the CDVAE baseline of −0.583 eV/atom to −0.449 (global search alone), −0.695 (plus symmetry-constrained local relaxation), −0.874 (plus primitive-cell reduction) and finally −1.027 eV/atom (plus dynamic search-space adaptation; Fig. 2d and Fig. S8), a cumulative 76% improvement. The same monotonic improvement holds on the Perovskite and OQMD (Open Quantum Materials Database)[36] test sets (Fig. S9, Table S3).

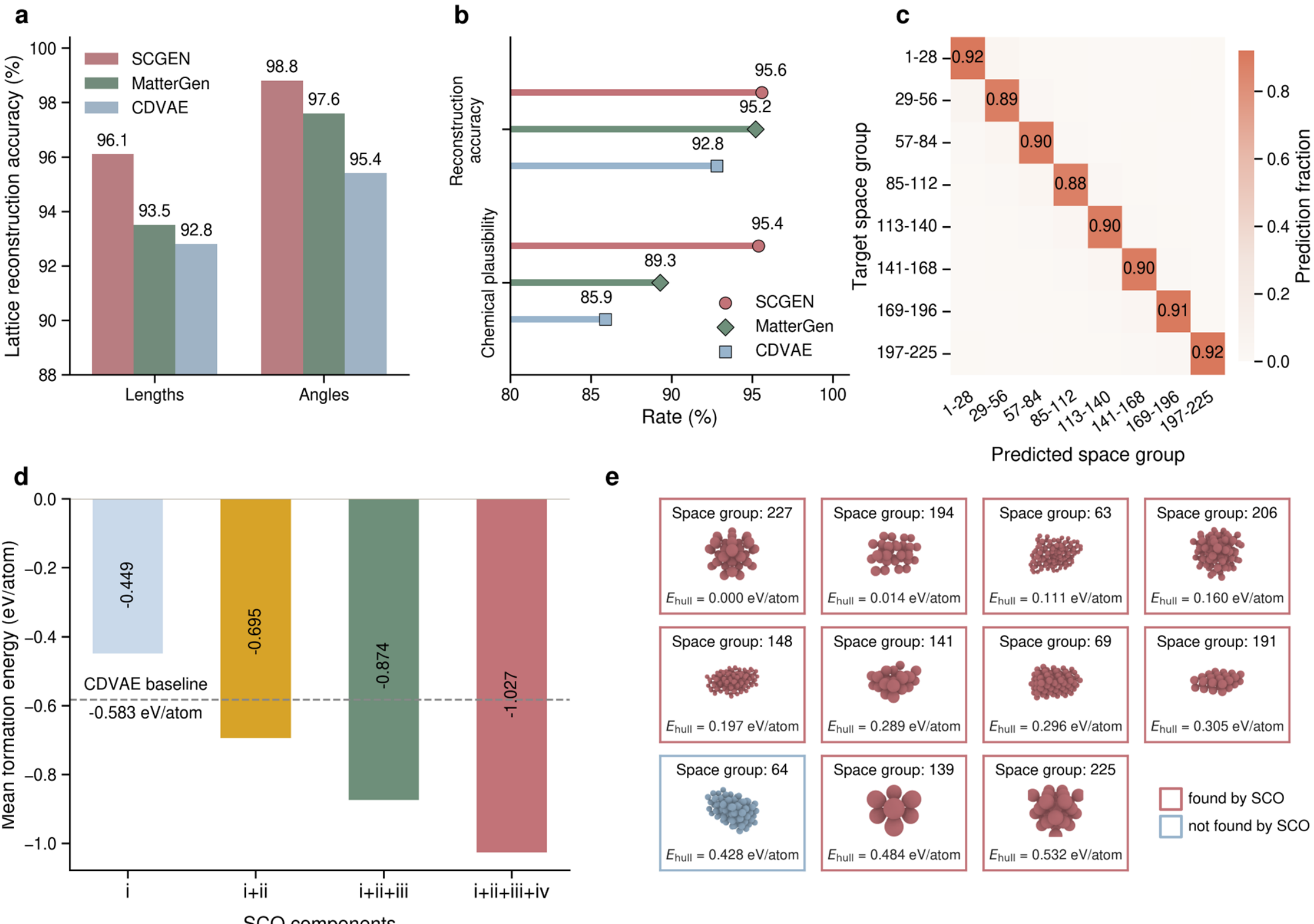

**Fig. 2. (a)** Lattice reconstruction accuracy for crystal-axis lengths and inter-axis angles, compared across SCGEN, MatterGen and CDVAE. **(b)** Composition reconstruction accuracy and chemical plausibility for the same three methods, shown as a Cleveland lollipop chart with 95% confidence intervals. **(c)** Confusion matrix for predicted vs target space group, binned into eight 28-group intervals spanning from 1 to 225; high values on the diagonal show SCGEN preserves the input space group across the full symmetry range. **(d)** Ablation of the four SCO components (i, i+ii, i+ii+iii, i+ii+iii+iv): the mean DFT formation energy of generated structures decreases monotonically as components are added, ending below the CDVAE baseline of −0.583 eV/atom (dashed line). **(e)** Reconstruction of 11 experimentally characterized Si polymorphs from the Materials Project: each card shows the rendered crystal structure with its space group and energy above hull; cards bordered in red are recovered by the SCO search, while blue-bordered cards are missed under the current settings.

The most stringent test of a physics-grounded search is whether it rediscovers the structures that are experimentally observed in nature. We therefore applied SCO to reconstruct elemental silicon by sweeping every space group and lattice compatible with the Si stoichiometry. Whereas a purely data-driven generator returns only what its training distribution makes probable, SCO directly navigates the true UMLP energy landscape. It successfully recovers 10 of the 11 experimentally characterized silicon polymorphs in their correct space groups (Fig. 2e), spanning the ground-state diamond phase as well as the high-pressure metastable forms that conventional generators systematically miss. SCGEN thus reproduces the structures dictated by the underlying physics, matching dedicated structure-search engines on their own ground while retaining the open-ended chemical novelty supplied by CIVAE.

**2.2 Overall performance benchmarking**

Current crystal-structure generators are mostly trained and evaluated on databases of structures with fewer than 20 atoms per unit cell (*i.e.*, MP-20[15]), limiting their applicability to the complex materials systems in practical discovery. To benchmark SCGEN against state-of-the-art generators on a substantially harder distribution, we assembled a unified corpus of 1,971,023 unique inorganic crystals by deduplicating across seven major repositories: materials project (MP)[37], open quantum materials database (OQMD)[36], crystallography open database (COD)[38], joint automated repository for various integrated simulations (JARVIS)[39], inorganic crystal structure database (ICSD)[40], open materials 2024 (OMat24)[41] and Alexandria[42]. This dataset spans the full complexity range of known inorganic materials and is, to the best of our knowledge, the largest reference set on which any open-source crystal generator has been benchmarked. We held out 10% for evaluation.

We systematically compared SCGEN with current advanced generative models (CDVAE[15], DiffCSP++[16], and MatterGen[17]) using this comprehensive database. Each model generated 5,000 candidates for subsequent DFT-based evaluation (Table 1). The decoupled paradigm embeds stability directly into the generation process itself. Because every structure is explicitly relaxed onto an energy minimum before it is reported, the generated population sits close to the convex hull by construction rather than by chance, and the benchmark confirms this expectation. SCGEN gives the lowest energy above hull (0.021 eV/atom), nearly halving that of the best generative baseline MatterGen (0.036 eV/atom) and improving on DiffCSP++ (0.195 eV/atom) and CDVAE (0.260 eV/atom) by roughly an order of magnitude (Fig. 3a).

**Table 1.** Comprehensive benchmarking of crystal structure generative models on the integrated database. "Inherent" indicates native capability without additional training, "Fine-tune" requires task-specific retraining, and "\" denotes functionality not available.

| Model | Stability (eV/atom) | Speed (Stru./hour) | Specified composition | Specified space group | Specified lattice | Model size |
|---|---|---|---|---|---|---|
| CDVAE | 0.260 | 1000 | / | / | / | 10.97M |
| DiffCSP++ | 0.195 | 1000 | / | Inherent | / | 12.28M |
| MatterGen | 0.036 | 900 | Fine-tune | Fine-tune | Fine-tune | 44.56M |
| **SCGEN** | **0.021** | **500** | **Inherent** | **Inherent** | **Inherent** | 11.16M |

**Fig. 3. (a)** Box-plot distribution of DFT-calculated energy above hull for structures generated by SCGEN, MatterGen, DiffCSP++ and CDVAE. **(b)** Novelty rate: percentage of generated structures whose composition is absent from the training set. **(c)** Dynamical-stability assessment: the left strip shows the minimum imaginary phonon frequency of unstable structures and the right bar report the count of dynamically stable structures (14 SCGEN vs 2 CDVAE out of 25). **(d,e)** Energy convex hulls for the Cu–Al binary system: formation energy versus copper fraction for 1,000 structures generated by **(d)** CDVAE and **(e)** SCGEN, with the constructed hull overlaid. **(f)** Histogram of generated space-group numbers, with inset pie charts of the P1 versus higher-symmetry split for SCGEN and CDVAE. **(g)** Conditional generation success rates for specified composition, space group, lattice parameters, and the joint specification of all three, compared across SCGEN, MatterGen and CDVAE. **(h)** Multi-axis radar chart summarising stability, symmetry preservation, validity, generation speed, controllability, and novelty across the four methods.

Importantly, this gain in stability does not compromise novelty. In SCGEN, novelty is ensured by CIVAE, which freely samples unseen compositions, while stability is the responsibility of SCO, which optimizes only the atomic coordinates and strictly preserves the chemistry. Because the two objectives are handled separately rather than traded off against each other, 78.0% of SCGEN structures carry compositions beyond the training database, remaining competitive with MatterGen (86.2%), DiffCSP++ (83.1%) and CDVAE (83.9%) (Fig. 3b).

We further assessed dynamical stability by computing phonon spectra for 25 randomly selected structures from each model. Of these, 14 SCGEN structures showed exclusively real phonon frequencies, against only 2 from CDVAE (Fig. 3c). A fully real phonon spectrum is the physical signature of a true energy minimum: because the hierarchical SCO relaxes every candidate on the UMLP energy landscape under symmetry, it settles into genuine minima, whereas direct coordinate prediction often halts at saddle points on the potential energy surface.

To illustrate the practical impact of these stability differences, we examined the Cu–Al binary system, generating 1,000 structures each with SCGEN and CDVAE. The resulting energy convex hulls differ markedly (Fig. 3d,e): SCGEN structures cluster near the thermodynamic stability line, with several compositions on or approaching the hull, whereas CDVAE structures scatter broadly above it. This underscores the capacity of SCGEN to generate synthesizable candidate materials across compositional space.

The same architectural distinction underpins the pronounced differences in symmetry preservation across methods. In SCGEN, the space group and its associated Wyckoff positions are imposed as hard constraints throughout coordinate determination rather than learned from data, ensuring that non-trivial symmetry is preserved by construction; consequently, 94.9% of the generated structures adopt space groups higher than P1. In contrast, models that generate atomic coordinates independently provide no such guarantee: uncorrelated perturbations violate the underlying symmetry operations and drive the structures toward the trivial P1 group. This is evident for CDVAE, where only 26.5% of structures retain any higher symmetry (Fig. 3f).

Beyond stability and symmetry, the decoupled architecture confers a further inherent advantage: controllability. Because SCGEN separates CI generation from coordinate determination, the CI is an explicit, editable input to SCO, so a user-specified composition, space group or lattice simply replaces the corresponding CIVAE output—a built-in guarantee rather than a trained capability. As a result, any single constraint, or any combination of the three, is satisfied with a 100% success rate without any fine-tuning (Fig. 3g). By contrast, conventional coordinate generators must be retrained for each target: MatterGen, even after per-condition fine-tuning, reaches only 89%, 79%, 73% and 58% for composition, space group, lattice and their joint constraint, while CDVAE performs worse (81%, 51%, 47% and 32%).

Having established individual advantages in stability, symmetry preservation and controllability, we consolidate all six evaluation metrics—stability, symmetry preservation, composition validity, generation speed, controllability and novelty—in a multi-dimensional radar comparison (Fig. 3h). SCGEN achieves the best overall balance among the evaluated methods, leading in five of the six dimensions. Notably, this performance is achieved with a remarkably compact network: SCGEN uses only 11.16M parameters (Table 1), roughly a quarter of the 44.56M in MatterGen, so that both the CIVAE and the SCO run on a laptop CPU without dedicated GPU resources. The sole exception is generation speed, a limitation that reflects a tunable design trade-off rather than a fundamental shortcoming of the decoupled paradigm. As quantified in Table 1, SCGEN generates approximately 500 structures per hour against 900–1,000 for purely generative approaches. This difference arises from the iterative SCO optimization stage despite extensive GPU-accelerated parallelization of the Wyckoff enumeration, global search and local relaxation.

This speed cost reflects the natural consequence of grounding generation in physics, since each structure is genuinely relaxed on an energy landscape rather than emitted in a single forward pass. The resulting speed–stability trade-off is continuously adjustable to the application: reducing the number of SCO optimization steps recovers speed comparable to competing methods while retaining 100% controllability, superior symmetry preservation and higher chemical plausibility with some reduction in structural stability. Rapid screening may favor this regime, whereas the discovery of synthesizable candidates benefits from full convergence. Taken together, these results show that stability, symmetry preservation and controllability are not independent metrics optimized in isolation, but common consequences of decoupling CI from coordinate determination. SCGEN thereby combines state-of-the-art quality with built-in controllability and the flexibility to balance computational cost against physical reliability as the application demands.

### 2.3 Application in the inverse design of PWS

The benchmarking in Section 2.2 establishes that SCGEN generates stable, symmetry-rich and fully controllable crystals, but these properties are valuable only if they accelerate the discovery of useful materials. We therefore deployed SCGEN within a complete inverse-design campaign aimed at one of the most demanding targets in materials chemistry: photocatalytic water-splitting (PWS) catalysts. A practical photocatalyst must satisfy three physical requirements simultaneously: correct band-edge alignment, thermodynamic stability, and resistance to electrochemical corrosion in water. Their conjunction, not any single criterion, makes the search space sparse. Because SCGEN relaxes every candidate onto a genuine energy minimum before screening, the optimizer searches almost exclusively among physically realizable materials. We therefore integrated SCGEN as the generative core of our previously developed Materials Generation and Evolutionary Crystal Search (MAGECS)[6] framework for the inverse design of PWS catalysts.

The first requirement governs the optical and electronic response. The conduction- and valence-band energies $E_{CB}$, $E_{VB}$ must straddle the $H^+/H_2$ and $H_2O/O_2$ redox potentials: $E_{CB}$ below the $H^+/H_2$ reduction potential (0 V vs. NHE, pH = 0) and $E_{VB}$ above the $H_2O/O_2$ oxidation potential (1.23 V)[43]. The decisive constraint is therefore not a sufficiently large gap ($E_g$ > 1.23 eV) alone, but simultaneous straddling of both redox levels. We trained a supervised GNN to predict band gaps and derived $E_{CB}$ and $E_{VB}$ from the geometric mean of atomic electronegativities:

$$E_{CB} = \sqrt[N]{\chi_1^{n_1} \chi_2^{n_1} \cdots \chi_k^{n_k}} - E_0 - \frac{1}{2} E_g \tag{1}$$

$$E_{VB} = E_{CB} + E_g \tag{2}$$

where $E_g$ and $E_0$ are the band gap and the free electron energy on the hydrogen scale (4.44 eV), respectively. The $\sqrt[N]{\chi_1^{n_1} \chi_2^{n_1} \cdots \chi_k^{n_k}}$ is the geometric mean of the electronegativities of all atoms in a structure, and the $\chi_k^{n_k}$ represents the product of electronegativities of species $k$ with $n_k$ atoms.

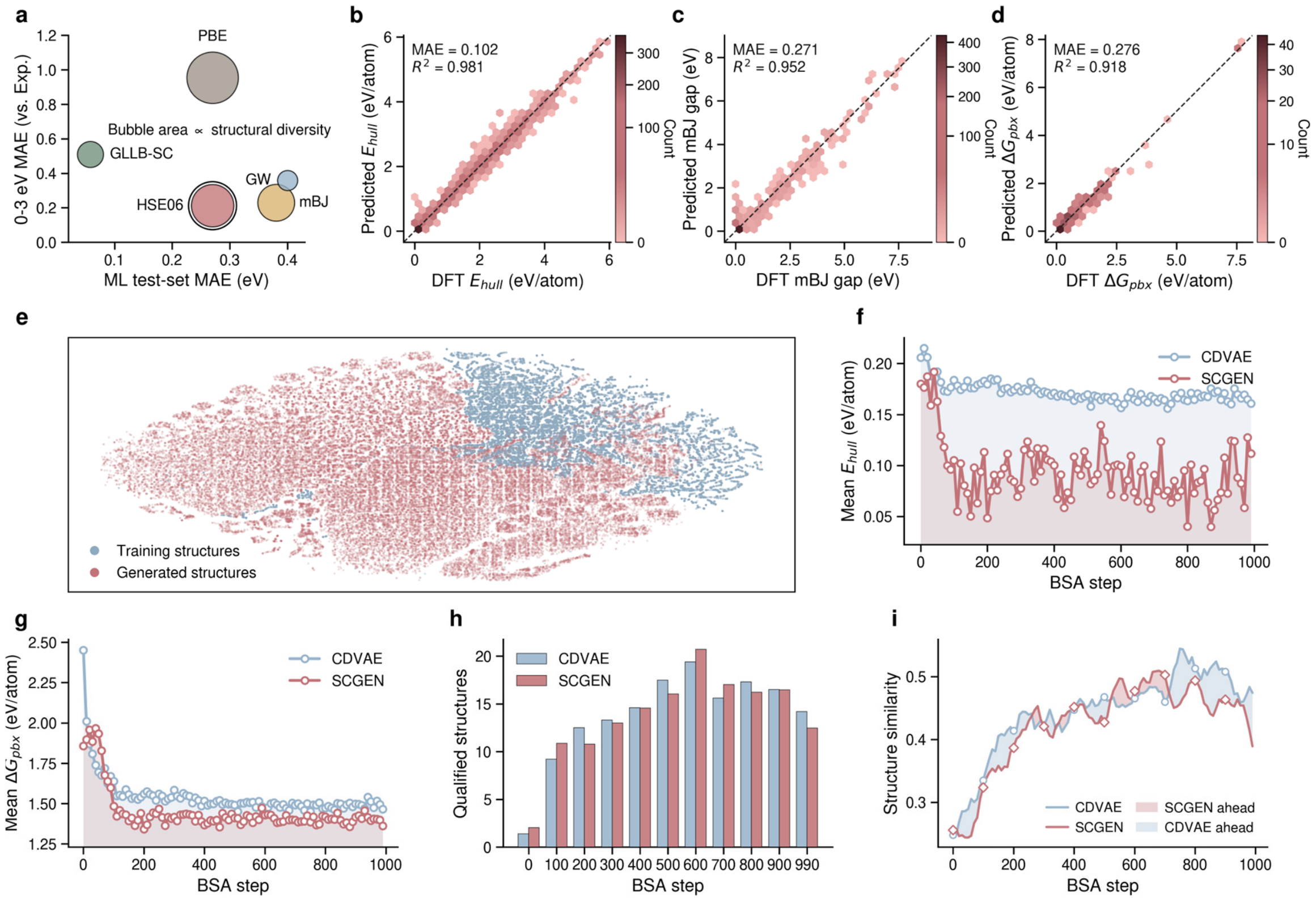

**Fig. 4. (a)** Band-gap computational method comparison: trade-off between ML-model predictive MAE (x-axis) and deviation from experimental values (y-axis), with bubble area proportional to structural diversity of training data; mBJ offers the optimal balance of accuracy, computational cost, and data diversity. **(b–d)** Property-predictor validation on held-out test sets: DFT-calculated versus GNN-predicted **(b)** $E_{hull}$, **(c)** mBJ band gap, and **(d)** $\Delta G_{pbx}$, with MAE and $R^2$ annotated and hexbin density along the diagonal. **(e)** t-SNE visualization of SCGEN-generated structures (red) overlaid on the training set (blue), showing broad coverage of the structural landscape with substantial exploration beyond training-set boundaries. **(f,g)** BSA optimization trajectories: running-average **(f)** $E_{hull}$ and **(g)** $\Delta G_{pbx}$ for SCGEN (red) and CDVAE (grey), with gradient fill under each curve. **(h)** Number of structures meeting the PWS band-gap criteria at each BSA step, shown as grouped bars (SCGEN red, CDVAE blue). **(i)** Inter-generation structural similarity over BSA steps, drawn as a smoothed two-model advantage ribbon — the area between the curves is shaded by the leading model at each step.

Selecting an appropriate band-gap calculation method is a balance between computational cost, predictive accuracy and the diversity of training data available. We compared mBJ (modified Becke-Johnson)[44], GLLB-SC (Gritsenko, Leeuwen, Lenthe, and Baerends-Solid State)[45], HSE (Heyd-Scuseria-Ernzerhof)[46] and GW[47] against experimental band gaps using literature benchmarks[48,49] (Fig. 4a, Fig. S10, Table S4). Restricted to the 0–3 eV semiconductor range relevant to PWS, the four methods give MAEs that differ by less than 0.08 eV, but mBJ requires roughly an order of magnitude less computation than GW or HSE. To choose the band-gap source for training the property predictor, we further evaluated these methods by the size and chemical diversity of their available training corpora (Note S5, Tables S5, S6). mBJ $E_g$ ranked second on a normalized diversity score, matched the test-set MAE of PBE $E_g$, and deviated less from experiment than PBE. It is therefore the optimal compromise for semiconductor-relevant ML prediction and the band-gap source we adopt throughout (Fig. 4a; selection criteria in Supplementary Note S5).

Beyond the band-gap criterion, candidate stability is decisive for PWS. We assess thermodynamic stability through $\Delta E_{hull}$, the energy above the convex hull in composition space, which represents the energy required for a material to decompose into the most stable phases of the same composition[50]. In addition to thermodynamic stability, long-term aqueous stability under oxidizing conditions (typically above 1.23 V vs RHE) is also essential for PWS[43], since many materials degrade in such environments. Our screening therefore emphasizes electrochemical stability through $\Delta G_{pbx}$, the Gibbs free energy of decomposition into Pourbaix-stable phases at a given pH and potential[51]. This metric captures whether a material remains stable in water, inherently or through a passivating layer[52], that is often overlooked in machine-learning studies. To bypass the expensive calculations of materials stability, especially the $\Delta G_{pbx}$, we trained another two GNNs to predict the $\Delta E_{hull}$ and $\Delta G_{pbx}$. The training data for $\Delta E_{hull}$ and $\Delta G_{pbx}$ are extracted from the JARVIS[39] and MP database[37], respectively.

After determining the target properties for PWS, we utilized SCGEN pretrained on the ~2 million structure database (Section 2.2) to create novel and diverse materials from latent vectors (embedding of material structures). To evaluate the target properties of generated materials for PWS, we implemented three pretrained GNN models to predict the $E_g$, $\Delta E_{hull}$, and $\Delta G_{pbx}$. Three CGTNets[53] (our previously developed GNN, details in Methods 3.2) for the prediction of $\Delta E_{hull}$ (Fig. 4b), mBJ $E_g$ (Fig. 4c), and $\Delta G_{pbx}$ (Fig. 4d) show excellent predictive accuracy ($R^2 > 0.9$) on testing data. The objective function combining the output of three CGTNets was then built:

$$S(\Delta E_{hull}) + S(\Delta G_{pbx}) + \varepsilon(E_{CB}, E_{VB}, E_g) \tag{3}$$

$$\varepsilon(E_{CB}, E_{VB}, E_g) = \begin{cases} 0 \ if \ E_{CB} \leqslant 0; E_{VB} \geqslant 1.23; 1.3 \leqslant E_g \leqslant 3.3 \\ 100 \ else \end{cases} \tag{4}$$

where the $S(\Delta E_{hull})$ and $S(\Delta G_{pbx})$ normalize $\Delta E_{hull}$ and $\Delta G_{pbx}$, respectively. The penalty term $\varepsilon(E_{CB,} E_{VB}, E_g)$ is zero when the $E_{VB}$, $E_{CB}$, $E_g$ of a generated structure meet the PWS standards and 100 otherwise. Accounting for the error of the mBJ $E_g$ prediction model (0.27 eV) and its deviation from experiment, the $E_g$ range in $\varepsilon(E_{CB,} E_{VB}, E_g)$ is set to 1.3-3.3 eV.

This objective function serves as the fitness of the bird swarm algorithm (BSA)[35,54], inspired by natural swarm intelligence (Fig. S11). The BSA processes batches of latent vectors and uses GNN-evaluation feedback to iteratively refine them until enough structures with optimal PWS properties are obtained, thereby combining generative modeling, property prediction and intelligent optimization (objective-value evolution in Fig. S12). SCGEN-generated structures explore substantially beyond the training distribution (t-SNE, Fig. 4e), and crucially, because SCO grounds every structure on the UMLP energy landscape, each fitness evaluation acts on a physically relaxed configuration rather than a raw network output, so the swarm navigates a smooth, physically meaningful property surface. This advantage is reflected in the running-average (every ten iterations) thermodynamic and aqueous stability relative to CDVAE (Fig. 4f,g), and in the number of structures meeting the PWS band-gap and band-edge criteria, which rose with each iteration to ~15-fold over the training set (Fig. 4h). These improvements stem mainly from the chemical-space exploration capability of BSA, so CDVAE and SCGEN perform comparably in this aspect; meanwhile, the structural similarity between consecutive generations remained around 50% (Fig. 4i), indicating continuous generation of novel structures beyond local optima in property space. Finally, BSA iterations also recovered several experimentally validated PWS structures, such as $SrTiO_3$, ZnO and $TiO_2$ (Fig. S13).

### 2.4 DFT validation and screening of inverse-designed PWS structures

We then apply four nested filters (Fig. 5a) to narrow the 200,000 SCGEN-generated candidates toward DFT-validated photocatalysts. The first two criteria, drawn from the literature, $\Delta E_{hull} \leq 0.1$ eV/atom and $\Delta G_{pbx} \leq 1.5$ eV/atom, ensure thermodynamic and aqueous stability respectively. To ease experimental synthesis, the third criterion removes structures with more than three species. In total, 3,318 structures met all four criteria and entered DFT calculations (elemental distribution in Fig. S14). DFT-calculated band gaps correlate strongly with CGTNet predictions (MAE 0.156 eV; Fig. 5b), confirming the reliability of CGTNet for band-gap prediction. Their distribution (Fig. 5c) shows that 95.6% of the DFT-calculated structures satisfy the PWS criteria, far above the training-set fraction, indicating that MAGECS guided the generator toward properties surpassing the training distribution.

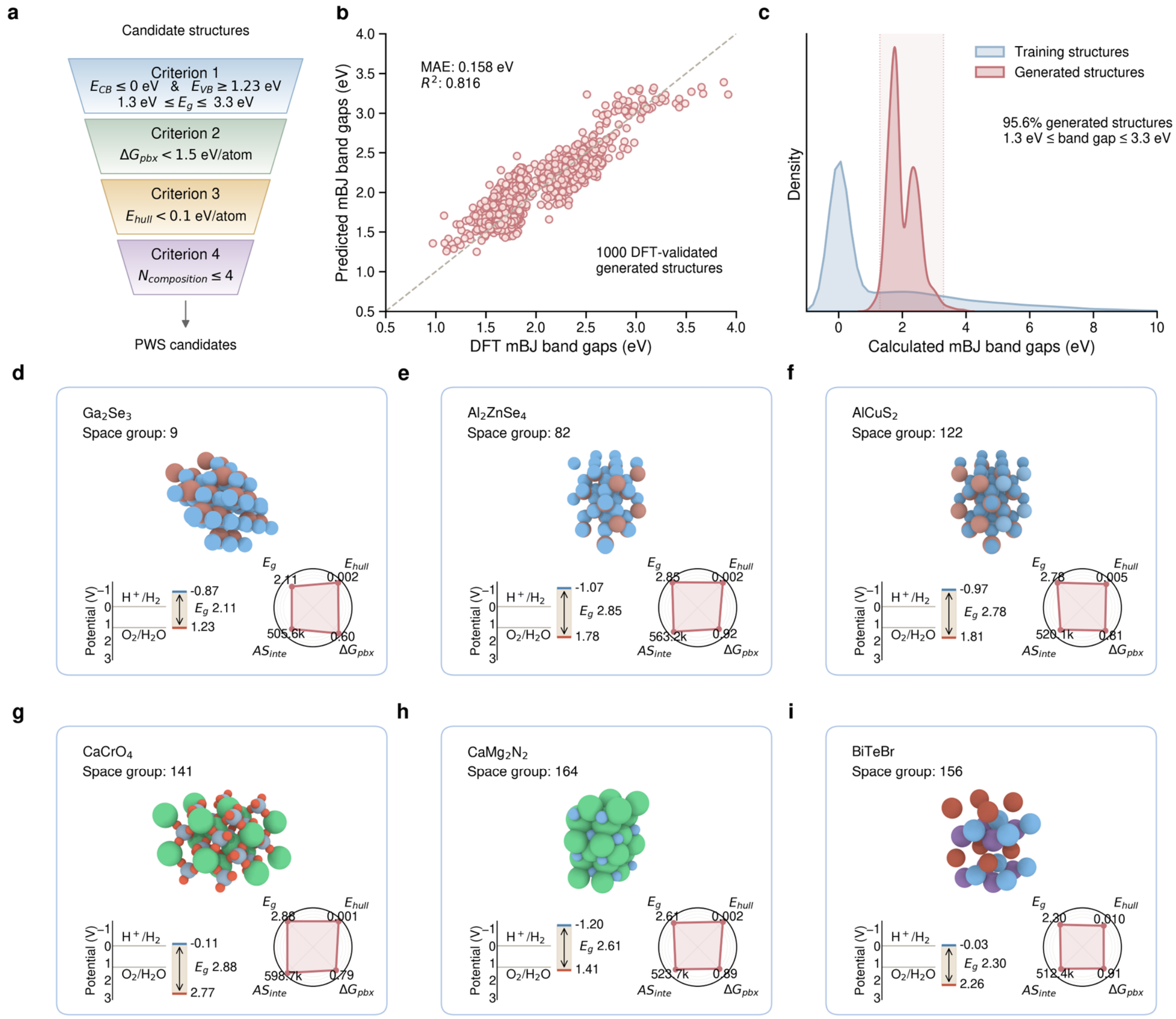

**Fig. 5. (a)** Four-criterion screening funnel for PWS photocatalyst candidates: (i) band-edge alignment $E_{CB} \leq 0$ eV and $E_{VB} \geq 1.23$ eV with visible-light gap $1.3 \leq E_g \leq 3.3$ eV, (ii) aqueous stability $\Delta G_{pbx} < 1.5$ eV/atom, (iii) thermodynamic stability $E_{hull} < 0.1$ eV/atom, and (iv) compositional simplicity $N_{composition} \leq 4$. **(b)** DFT-calculated versus GNN-predicted mBJ band gaps for generated structures, validating the reliability of the property predictor for screening. **(c)** Band-gap distribution of the training set (blue) and generated structures (red); generated structures are concentrated within the 1.3–3.3 eV window (shaded) required for photocatalytic water splitting, showing significant enrichment over the training distribution. **(d–i)** Six representative photocatalyst candidates: each panel shows the crystal structure (2×2×2 supercell), formula and space group, band-edge alignment diagram relative to the $H^+/H_2$ and $O_2/H_2O$ redox potentials, and a four-axis radar chart summarizing Ehull, ΔGpbx, ASinte, and Eg. Materials shown: **(d)** $Ga_2Se_3$, **(e)** $Al_2ZnSe_4$, **(f)** $AlCuS_2$, **(g)** $CaCrO_4$, **(h)** $CaMg_2N_2$, **(i)** BiTeBr.

Having narrowed the pool on thermodynamic and aqueous-stability grounds, we next assessed the dynamical and optical properties that ultimately govern photocatalytic performance. We further performed phonon calculations on candidates selected according to their DFT-calculated $\Delta E_{hull}$, $E_g$, $E_{CB}$, and $E_{VB}$. Since DFT values are available at this stage, the $E_g$ criterion was tightened to 1.6–2.4 eV. The absence of imaginary phonon frequencies confirms dynamical stability against lattice vibrations. We also computed integrated absorption spectra values ($AS_{inte}$) within the 1.6–2.4 eV range to evaluate light-harvesting capability, providing a measure of how effectively each candidate captures visible light.

**Table 2** Top 22 generated structures after screening by DFT band gaps and band edges, phonon spectral stability, $E_{hull}$, $\Delta G_{pbx}$, and optical absorption spectra. Calc. and Pred. are abbreviations for calculated and predicted, respectively.

| Material | $E_{hull}$ (eV/atom) | $\Delta G_{pbx}$ (eV/atom) | SP | $AS_{inte}$ | Calc. $E_g$ (eV) | Calc. $E_{cb}$ (V) | Calc. $E_{vb}$ (V) | Pred. (eV) | Pred. $E_{cb}$ (V) | Pred. $E_{vb}$ (V) |
|---|---|---|---|---|---|---|---|---|---|---|
| $Cs_4I_7$ | 0.001 | 0.045 | 10 | 645793 | 2.651 | -1.373 | 1.278 | 2.704 | -1.399 | 1.305 |
| $Al_2ZnSe$ | 0.002 | 0.915 | 82 | 563212 | 2.851 | -1.073 | 1.778 | 2.682 | -0.988 | 1.694 |
| $AlCuS_2$ | 0.005 | 0.805 | 122 | 520129 | 2.782 | -0.975 | 1.807 | 2.902 | -1.035 | 1.867 |
| $CaCrO_4$ | 0.001 | 0.789 | 141 | 598726 | 2.883 | -0.113 | 2.770 | 2.723 | -0.033 | 2.69 |
| $CaMg_2N$ | 0.002 | 0.891 | 164 | 523685 | 2.612 | -1.203 | 1.409 | 2.651 | -1.223 | 1.428 |
| CuI | 0.009 | 0.771 | 164 | 502664 | 2.545 | -0.211 | 2.334 | 2.812 | -0.344 | 2.468 |
| $Ga_2Se_3$ | 0.002 | 0.598 | 9 | 505612 | 2.106 | -0.872 | 1.234 | 2.159 | -0.898 | 1.261 |
| GaSe | 0.005 | 0.856 | 194 | 598613 | 2.651 | -1.415 | 1.236 | 2.778 | -1.479 | 1.299 |
| $Ge_3N_4$ | 0.001 | 0.689 | 176 | 556523 | 2.662 | -0.151 | 2.511 | 2.733 | -0.187 | 2.546 |
| MgSe | 0.001 | 0.865 | 186 | 541298 | 2.458 | -0.925 | 1.533 | 2.479 | -0.936 | 1.543 |
| ZnSe | 0.001 | 0.598 | 216 | 623149 | 2.255 | -0.309 | 1.946 | 2.274 | -0.319 | 1.955 |
| ZnTe | 0.001 | 0.645 | 186 | 594123 | 2.064 | -0.394 | 1.67 | 2.129 | -0.426 | 1.703 |
| $TlSnO_2$ | 0.001 | 0.63 | 59 | 546931 | 2.232 | -0.272 | 1.96 | 2.159 | -0.236 | 1.923 |
| $Cs_2SnO_3$ | 0.001 | 0.73 | 59 | 534562 | 2.799 | -1.31 | 1.489 | 2.852 | -1.337 | 1.515 |
| $Tl_5Sn_4O_1$ | 0.001 | 0.708 | 3 | 498526 | 2.326 | -0.329 | 2.326 | 2.364 | -0.366 | 2.364 |
| $Nd_3Cu_2$ | 0.002 | 0.804 | 47 | 516647 | 2.256 | -0.071 | 2.185 | 2.137 | -0.012 | 2.125 |
| CuCl | 0.011 | 0.778 | 160 | 498564 | 2.733 | -0.288 | 2.445 | 2.798 | -0.256 | 2.542 |
| $SrPbO_2$ | 0.075 | 0.87 | 51 | 644329 | 2.2 | -0.521 | 1.679 | 2.311 | -0.577 | 1.734 |
| $BeZnO_2$ | 0.076 | 1.044 | 12 | 632356 | 2.171 | -0.012 | 2.159 | 2.261 | -0.059 | 2.204 |
| $SrCdO_2$ | 0.063 | 1.129 | 8 | 569508 | 1.703 | -0.087 | 1.617 | 1.636 | -0.053 | 1.583 |
| $BaSnO_2$ | 0.089 | 1.198 | 51 | 451265 | 1.901 | -0.354 | 1.548 | 1.789 | -0.297 | 1.491 |
| BiTeBr | 0.01 | 0.91 | 156 | 512358 | 2.297 | -0.034 | 2.263 | 2.256 | -0.013 | 2.243 |

While thermodynamic, dynamical and aqueous stability are necessary conditions for PWS materials, they are insufficient to guarantee experimental realizability. To bridge this gap between computational prediction and practical synthesis, we employed our previously developed crystal synthesis large language models (CSLLM) framework[55]. Rather than relying solely on energy-based descriptors, CSLLM uses fine-tuned large language models to accurately predict synthesizability from crystal-structure representations[55]. From the 3,318 DFT-validated structures, CSLLM identified 1,260 candidates as potentially synthesizable, substantially narrowing the search space. Through this hierarchical pipeline—combining thermodynamic and aqueous stability, dynamical stability, optical properties, and synthesizability—we finally identified the top 22 synthesizable candidates, whose full screening metrics are summarized in Table 2. For these 22 structures, CSLLM further suggested viable synthetic routes and precursor combinations, with the recommended synthesis methods and recipes tabulated in Table S9.

The six representative candidates in Fig. 5d–i span three chemical families—the chalcogenides $Ga_2Se_3$, $Al_2ZnSe_4$ and $AlCuS_2$, the ternary nitride $CaMg_2N_2$, the oxide $CaCrO_4$ and the layered halide chalcogenide BiTeBr—and four crystal systems, illustrating the chemical and structural breadth SCGEN achieves under photocatalytic constraints. All six lie within 5 meV/atom of the convex hull ($\Delta E_{hull}$ = 0.001–0.010 eV/atom) and well below the aqueous-stability threshold ($\Delta G_{pbx} \leq 0.92$ eV/atom), meeting the thermodynamic and electrochemical requirements for photocatalytic water splitting (PWS). The chalcogenides $Ga_2Se_3$ ($E_g$ = 2.11 eV) and $AlCuS_2$ ($E_g$ = 2.78 eV) have conduction-band edges 0.87 and 0.98 eV below the $H^+/H_2$ potential, giving ample overpotential for hydrogen evolution while preserving visible-light absorption. $Al_2ZnSe_4$ ($E_g$ = 2.85 eV) extends the family into wider-gap territory suited to UV–blue overall splitting. $CaMg_2N_2$ ($E_g$ = 2.61 eV) is a quasi-layered P-3m1 nitride whose valence band of 1.41 V is set by Mg–N hybridization, an alignment that has made related ternary nitrides hard to engineer by composition alone, yet which SCGEN finds without supervision. $CaCrO_4$ is the only oxide among the six. Its near-zero conduction-band edge ($E_{CB} = -0.11$ V) places hydrogen evolution at marginal overpotential, but its large valence-band offset ($E_{VB}$ = 2.77 V) and integrated absorption strength ($AS_{inte} = 5.99 \times 10^5$) point to strong oxygen-evolution performance and visible absorption from the Cr-3*d*–O-2*p* valence band. BiTeBr is a known

Rashba semiconductor usually studied for spin–orbit physics rather than photocatalysis. Its rediscovery here ($E_g$ = 2.30 eV, with a near-perfect band-edge straddle of 0 and 1.23 V) shows that SCGEN explores beyond the training distribution rather than memorizing it, and points to an experimental direction so far overlooked.

## Conclusion

In conclusion, SCGEN reframes generative crystal design as a physics-grounded union of two complementary paradigms. It inherits chemical novelty from generative learning, sampling composition, space group and lattice from a CIVAE-learned distribution, and inherits structural quality from crystal-structure prediction, locating atomic positions through symmetry- and Wyckoff-constrained optimization on a UMLP energy landscape. Benchmarked on 1,971,023 inorganic crystals, SCGEN achieves a state-of-the-art mean energy above hull of 0.021 eV/atom with comparable novelty, a 42% reduction relative to MatterGen, while satisfying any composition, space-group, or lattice constraint with 100% success and no task-specific retraining. Applied to photocatalytic water splitting, it delivers 22 DFT-confirmed candidates across four chemical families.

These achievements stem from three architectural advantages. First, the physics-grounded formulation encodes stability and symmetry into generation by construction, ensuring that every output relaxes to a genuine energy minimum at its prescribed Wyckoff positions. Second, crystallography-aware optimization restricts the search to the symmetry-permitted degrees of freedom, contracting the configurational space by orders of magnitude and rendering coordinate determination tractable at database scale. Third, because CIVAE and SCO interface solely through CI, every CI constraint is controllable by construction, and either component may be upgraded independently as better generative architecture or UMLPs appear. By grounding generation in physics and making it computationally efficient, SCGEN offers a new paradigm for inverse materials design, demonstrating that stable, novel, and controllable crystal-structure generation does not require generating atomic coordinates directly.

## 3. Methods

### 3.1. SCGEN

For the initial CIVAE testing, we selected 75,000 crystal structures with no more than 40 atoms from the OQMD database for training, with the training and validation sets divided in an 8:2 ratio. In the SCO section, our testing employed the bird swarm algorithm for global optimization and the conjugate gradient method for local optimization, with the universal machine learning potential MatterSim used to calculate the optimization objective function.

### 3.2. CGTNet

For each of the three properties, we trained our previously proposed transformer-based graph neural network model, CGTNet. This model adopts a transformer framework and integrates bond length and bond angle information. This model has demonstrated better accuracy and small data friendliness compared to CGCNN, SchNet, PaiNN, DimeNet++, and GemNet. The main hyperparameters used for training CGTNets are shown in Table S7.

### 3.3. Bird swarm algorithm

The BSA algorithm is inspired by the swarm intelligence observed in bird swarms. Birds exhibit three main behaviors: foraging, vigilance, and flight. These social interactions help birds find food and avoid predators, increasing their chances of survival. BSA models these behaviors with five simplified rules, giving the algorithm excellent optimization efficiency and the ability to escape local optima. Thus, the BSA is used for global exploration of the chemical space in our framework. The hyperparameters used for the BSA optimization are listed in Table S8.

### 3.4. Automated DFT calculations

All the DFT energies in this work were calculated by the VASP package[56]. For these computations, a plane wave basis set with a cutoff energy of 520 eV and a k-point mesh of 1000/number of atoms in the cell was utilized to ensure accuracy. The same convergence criteria as established for other DFT calculations were maintained, with the energy convergence set at $5 \times 10^{-5}$ eV. The mBJ exchange-correlation functional, known for its effectiveness in band gap estimation, was employed in these calculations.

## Acknowledgements

This work was supported by the National Key Research and Development Program of China (grant 2021YFA1500700), the Natural Science Foundation of China (grant 22033002, 22373013, T2321002), and the Basic Research Program of Jiangsu Province (BK20232012, BK20222007), Jiangsu Provincial Scientific Research Center of Applied Mathematics (BK20233002) and the Fundamental Research Funds for the Central Universities. We thank the Big Data Computing Center of Southeast University for providing the facility support on the calculations.